\documentclass[12pt]{amsart}

\usepackage{etoolbox}
\usepackage{amsthm}
\usepackage{enumitem}
\usepackage{tikz}
\usepackage{booktabs}
\usepackage{url}
\usepackage{amssymb,amsmath,graphicx}
\expandafter\def\expandafter\UrlBreaks\expandafter{\UrlBreaks
  \do\-\do\/}

\usepackage[margin=1in,letterpaper,portrait]{geometry}
\usetikzlibrary{arrows,shapes,automata,backgrounds,decorations,petri,positioning}
\usetikzlibrary{decorations.pathreplacing,angles,quotes}
\usetikzlibrary[calc,intersections,through,backgrounds,arrows,decorations.pathmorphing]
\tikzstyle{edge} = [fill,opacity=.5,fill opacity=.5,line cap=round, line join=round, line width=50pt]

\theoremstyle{plain}
\theoremstyle{definition}
\newtheorem*{definition-nonum}{Definition}

\newcommand{\hsepline}[1]{ 
\draw (#1)++(0,-.3,0) --++(0,1.6);}

\newcommand{\honebar}[2]{ 
\draw[fill=#1-col] (#1) rectangle ++(#2,1);
\draw (#1) ++(#2,0) coordinate (#1);}

\newcommand{\htwobar}[3]{ 
\draw[fill=#2-col] (#2) rectangle ++(#3,.5);
\draw[fill=#1-col] (#2)++(0,.5) rectangle ++(#3,.5);
\draw (#2) ++(#3,0) coordinate (#2);}

\newcommand{\hthreebar}[4]{ 
\draw[fill=#3-col] (#3) rectangle ++(#4,.333);
\draw[fill=#2-col] (#3)++(0,.333) rectangle ++(#4,.333);
\draw[fill=#1-col] (#3)++(0,.666) rectangle ++(#4,.334);
\draw (#3) ++(#4,0) coordinate (#3);}

\newcommand{\hfourbar}[5]{ 
\draw[fill=#4-col] (#4) rectangle ++(#5,.25);
\draw[fill=#3-col] (#4)++(0,.25) rectangle ++(#5,.25);
\draw[fill=#2-col] (#4)++(0,.5) rectangle ++(#5,.25);
\draw[fill=#1-col] (#4)++(0,.75) rectangle ++(#5,.25);
\draw (#4) ++(#5,0) coordinate (#4);}

\newcommand{\hfivebar}[6]{ 
\draw[fill=#5-col] (#5) rectangle ++(#6,.2);
\draw[fill=#4-col] (#5)++(0,.2) rectangle ++(#6,.2);
\draw[fill=#3-col] (#5)++(0,.4) rectangle ++(#6,.2);
\draw[fill=#2-col] (#5)++(0,.6) rectangle ++(#6,.2);
\draw[fill=#1-col] (#5)++(0,.8) rectangle ++(#6,.2);
\draw (#5) ++(#6,0) coordinate (#5);}

\definecolor{hoar-col}{rgb}{0.902,0.380,0.004}
\definecolor{bond-col}{rgb}{0.992,0.722,0.388} 
\definecolor{poliquin-col}{rgb}{0.698,0.671,0.824} 
\definecolor{golden-col}{rgb}{0.369,0.235,0.600} 
\definecolor{nytimes-golden}{rgb}{0.133,0.498,0.764}

\begin{document}

\title{Accumulation charts for instant-runoff elections}

\author{Bridget Eileen Tenner}
\address{Department of Mathematical Sciences, DePaul University, Chicago, IL, USA}
\email{bridget@math.depaul.edu}
\thanks{Research partially supported by Simons Foundation Collaboration Grant for Mathematicians 277603.}

\author{Gregory S.~Warrington}
\address{Department of Mathematics \& Statistics, University of Vermont, Burlington, VT, USA}
\email{gregory.warrington@uvm.edu}
\thanks{Research partially supported by Simons Foundation Collaboration Grant for Mathematicians 429570.}

\maketitle

Who won the election? How did it play out? 
When election results are announced and analyzed, details matter. 
A standard bar chart gets the job done for a
plurality election, but it falls short for \emph{instant-runoff voting (IRV)} elections, in which candidates
are eliminated one-by-one according to voters' rankings. A single bar chart or table cannot capture the
multiple rounds of IRV tallying, thus obscuring and perhaps undermining the public's trust in IRV itself.

The media frequently report IRV results using multiple tables or bar
charts, in a (not altogether successful) attempt to communicate the
full story of an election.  While IRV has been used for decades in
state and local elections in the U.S.~\cite{fairvote}, the recent
adoption of IRV by the state of Maine for a number of state races,
including the U.S. Senate, marked the first time that such an
alternative to plurality voting would be used in a U.S. congressional
election~\cite{maine-first}. Figure~\ref{fig:nyt maine results}
illustrates a typical way in which the results of the 2018 election
for Maine's 2nd Congressional District were communicated.
\begin{figure}[htbp]
\begin{tabular}{lrc}\toprule
  Candidate & Votes & Ranked-Choice Votes\\\midrule
Golden (winner) & 131,954 & 139,231 \\
Poliquin        & 134,061 & 136,326 \\
Bond            & 16,452  & ---     \\
Hoar            & 6,865   & ---     \\\bottomrule
\end{tabular}
 \caption{Essential data conveyed by \textit{The New York
     Times}~\cite{nyt-maine} and \textit{Wikipedia}~\cite{wiki} showing the
   2018 election results from Maine's 2nd Congressional District. (See
   also \textit{The Washington Post} and \textit{Politico}~\cite{wapo,politico}, which omit
   the third column, and \textit{Ballotpedia}~\cite{ballot}, which omits the
   second column.)}
\label{fig:nyt maine results}
\end{figure}

Unfortunately, this represents the IRV election as a plurality
election plus some mysterious extra steps that were invoked when no
candidate won a majority of the valid first-round votes, retaining
little connection to the nature and features of IRV.

One of the selling
points of IRV is the ability of a voter, without wasting her vote, to vote for a candidate who will likely receive a small number of first-preference ballots: such a candidate will be eliminated in an
early round and each ballot reallocated to the highest-ranked candidate who is still in the running. The fact that IRV allows voters to rank candidates, and then processes those rankings, has an indisputable effect on how people vote in an IRV election. 

We define a candidate's \emph{coalition} of support in an IRV election as the collection of voters whose ballots are allocated to that candidate during some round of IRV. 
The first-round tally in IRV reflects an
important part of voters' preferences and candidates' coalitions, but
the results of that first round do not tell the full story of the
election --- particularly for a candidate who receives
the most votes in the first round but does not ultimately win the
majority. For example, as Wright claimed after his loss in the 2009
mayoral election in Burlington, Vermont (conducted under IRV, see
Figure~\ref{fig:BFP}), ``under the plurality system, I would have won
tonight''~\cite{bfp-art}. That first-round tally is just one facet of
what the voters prefer. Moreover, the voters might not have voted for
those top choices if they had not been casting ranked-choice votes at
the time. The point of IRV is that important data is contained in the
lower-ranked votes as well. Just as the first-lap leader of a race has
no claim on being the eventual winner, neither is the leader after the
first round of IRV tallying anything more than a temporary front
runner.

Besides avoiding this emphasis on first-round results, we propose that an election graphic --- whatever the election framework --- should, at a minimum, achieve the
following (cf.~\cite{civicdesign}).

\begin{quote}
\noindent \textbf{Objectives of an election graphic:}
\begin{itemize}
\item Be easy to understand.
\item Clearly indicate a winner.
\item Reflect the methodology of the election procedure.
\item Summarize the ballots that were cast.
\end{itemize}
\end{quote}

The first objective is a precondition for any graphical depiction of
data. Likewise, the point of holding an election is to determine a
winner, justifying the second objective. Consumers of election results have different
priorities, and the last two
objectives combine to serve those needs. For
example, a voter might be most interested in how the winner was
determined from the votes cast, whereas a candidate might be most
interested in the coalitions of support that were demonstrated through those votes. These issues are especially
pertinent in an IRV election, when not all members of a candidate's
broadest coalition may have ranked that candidate as their top choice.

The third and fourth objectives are nontrivial in the context of IRV
elections, where the winner is determined by a
multi-step algorithm. As discussed above, while the first round of
this algorithm looks deceptively like a simple plurality election, its
role in the IRV framework is different. Furthermore, the voters' expressed
preferences are, themselves, complicated and hard to summarize.  A
list of votes according to frequency quickly becomes unwieldy. For
example, there are $325$ possible ranked-choice votes on five 
candidates if a voter is not required to rank all five options: there are $5!=120$ ways to rank all of the candidates, and $5!(1/1! + 1/2! + 1/3! + 1/4!) = 205$ rankings of between one and four candidates.

In this paper, we propose an \emph{accumulation chart} for
illustrating the results of IRV elections. It can be read at a glance,
the winner is clearly indicated, it shows the impact of each round of
the instant-runoff procedure, and every vote can be traced throughout
the tallying. Moreover, candidates can use the accumulation chart to understand their
coalitions of support. These coalitions are displayed up through a candidate's last
round of participation, with a full description of the ``pedigree'' of all
votes that were included in the candidate's final vote totals. 

Our proposal is organized as follows. In the section titled ``Ranked-choice and instant-runoff voting'', we define instant-runoff voting and highlight its main features. In
``Examples of IRV election graphics,'' we discuss how the results of recent IRV elections have been reported in the press and some of the questions left unanswered by that reporting. In ``Accumulation charts,'' we introduce these tools, demonstrate their value using those same recent IRV elections, and explore their benefits for different constituencies of interested parties. We conclude with a call to use accumulation charts more broadly.

\section*{Ranked-choice and instant-runoff voting}
\label{sec:irv}

Instant-runoff voting considers not only each voter's first-choice
candidate but, as necessary, her lower-ranked choices as well. Whereas
\emph{plurality} (a.k.a. \emph{first-past-the-post} or
\emph{winner-takes-all}) elections require each voter to name at most
one preferred candidate, IRV involves a more complicated vote.

\begin{definition-nonum}
  A \emph{ranked-choice vote} is a list of candidates, ordered from top choice to bottom.
\end{definition-nonum}

Different methods have been proposed for identifying a
winner from a collection of ranked-choice votes. Our focus is on the following procedure.

\begin{definition-nonum}
\emph{Instant-runoff voting (IRV)} (also termed \emph{ranked-choice voting} and \emph{alternative voting}~\cite{fairvote,electoral-systems}) is a method for tallying ranked-choice votes. It works, iteratively in \emph{rounds}, as follows.
\begin{quote}
\begin{enumerate}\renewcommand{\labelenumi}{Step \theenumi:}
\item The top-ranked selection of each vote is tallied.
\item If there are two candidates, then the candidate with the most votes is declared the winner.
\item If there are more than two candidates, then the candidate with the fewest votes is eliminated and that candidate's name is stricken from each vote, with lower-ranked candidates advancing one position in that vote's ranking. The procedure now begins again with a new round.
\end{enumerate}
\end{quote}
\end{definition-nonum}

We present the following example as a demonstration.

\begin{quote}
\textbf{Example.} Irvtown is electing a mayor from three candidates: Alf, Bugs, Chester. The election is being conducted under IRV. The 22 votes cast in the election are described in the table below:
\begin{center}
\begin{tabular}{c|l|l|l}\toprule
\textbf{\# votes} & \textbf{1st} & \textbf{2nd} & \textbf{3rd}\\\midrule
1 & Alf & Bugs & Chester\\
5 & Alf & Chester & Bugs\\
3 & Bugs & Alf & Chester\\
2 & Bugs & Chester & --- \\
4 & Bugs & Chester & Alf\\
3 & Chester & Alf & Bugs\\
4 & Chester & Bugs & Alf\\\bottomrule
\end{tabular}
\end{center}
For example, two voters ranked Bugs as their top choice, Chester as their second choice, and listed no one as their third choice.

After the first round of IRV tallying, Alf receives six votes, Bugs receives nine votes, and Chester receives seven votes. Alf, with the fewest votes, is eliminated. In the second round, one vote is reallocated from Alf to Bugs, and five votes are reallocated from Alf to Chester. Thus, after the second round of tallying, Bugs has received 10 votes and Chester has received 12 votes. Of these two remaining candidates, Chester is declared the winner.
\end{quote}

The iterative nature of IRV relies on something that does not exist in plurality elections, and which is, in fact, critical to understanding a candidate's coalition.

\begin{definition-nonum}
Suppose that Candidate A accumulates a particular ranked-choice vote in a round of IRV. The \emph{pedigree} of that accumulated vote is the ranked list of all candidates who appeared on that particular ballot ahead of and including Candidate A.
\end{definition-nonum}

Thus, in the preceding example, the pedigrees of the votes Chester accumulates by the final round record the votes of voters who preferred Chester to all other candidates, as well as voters who preferred Alf to all others, but thought Chester was preferable to Bugs.

IRV is frequently described with a Step 1.5: ``If any candidate has
received a majority of the votes tallied in a given round, then that
candidate is declared the winner.'' Including this step might
shorten the tallying process and it does not change the eventual outcome of the election. In practice, this rule appears to
be frequently followed (e.g., \cite[\S 4.2.A]{merules} and \cite{sfrules}).
However, we choose to omit this option to truncate the IRV procedure because, in our view, Step 1.5 shifts the emphasis of the IRV algorithm in an
unwelcome direction. It is valuable to know what occurs in all rounds of the IRV algorithm, including the full pedigrees of all ballots.

\section*{Examples of IRV election graphics}
\label{sec:nyt}

\begin{figure}[htbp]
  \includegraphics[width=.75\linewidth]{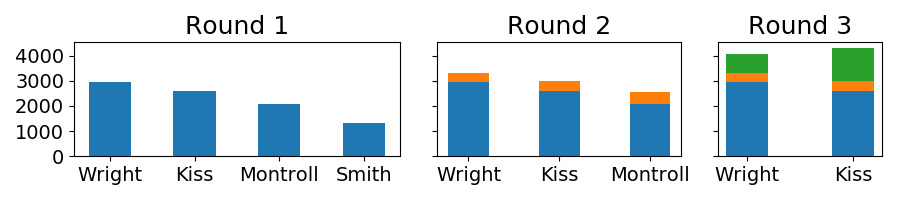}
\caption{Essential data conveyed by \emph{Burlington Free
    Press}~\cite{bfp-art} from the 2009 mayoral election in
  Burlington, Vermont.}\label{fig:BFP}
\end{figure}

Before introducing accumulation charts, we look at IRV election
results in the media. The Maine results shown in Figure~\ref{fig:nyt
  maine results} give the candidates' vote tallies after only two of
the rounds, suppressing earlier (including write-in candidates) and
intermediate (after Hoar's elimination) data. The 2009 mayoral
election in Burlington, Vermont was also conducted under IRV, and
Figure~\ref{fig:BFP} captures how the results of that election were
presented in the \emph{Burlington Free Press}~\cite{bfp-art} (see
also~\cite{mnsankey} for analogous reporting of the results of
Minneapolis's 2017 mayoral election).

Faced with either of these examples, a reader might well ask, which
chart should I look at? Which graphic's winner is the election's
winner?  Why did two candidates switch positions between the first and
second table? I ranked the candidates A$>$B$>$C$>$D --- where is my
vote in this picture? And so on. These graphics have suppressed so
much information that they barely reflect the IRV process, and they
give no sense of what votes were cast in the election. For example, of
the $767$ (green) votes added to Wright's tally
in Figure~\ref{fig:BFP}'s chart for Round 3, there is no way to
distinguish between votes that begin with the following rankings.

\begin{center}
\begin{tabular}{|l|}
\hline
1. Montroll\raisebox{5pt}[15pt][0pt]{\ } \\
2. Wright\raisebox{5pt}[0pt][10pt]{\ }\\
\hline
\multicolumn{1}{r}{\ }
\end{tabular} \ \ 
\begin{tabular}{|l|}
\hline
1. Montroll\raisebox{5pt}[15pt][0pt]{\ } \\
2. Smith\raisebox{5pt}[0pt][0pt]{\ }\\
3. Wright\raisebox{5pt}[0pt][10pt]{\ }\\
\hline
\end{tabular} \ \ 
\begin{tabular}{|l|}
\hline
1. Smith\raisebox{5pt}[15pt][0pt]{\ } \\
2. Montroll\raisebox{5pt}[0pt][0pt]{\ }\\
3. Wright\raisebox{5pt}[0pt][10pt]{\ }\\
\hline
\end{tabular}

\end{center}

Also, these graphics do a weak job of describing candidates' coalitions. The information gained
becomes more vague with each successive round, and there is ambiguity
about the coalitions forming support for the candidates who make it to
the final round. For example, Wright might
want to know the precise makeup of his support in order to craft a
subsequent campaign. Kiss, similarly, might want to know what
coalition put him over the top, and how best to represent them during
his time in office. Neither candidate would be able to find this
information from the data as currently depicted.

\section*{Accumulation charts}\label{sec:accumulation}

Accumulation charts are motivated by the fact that, in IRV, candidates can
accumulate votes during each round of the procedure. Illustrating
that accumulation makes substantial progress towards achieving the
objectives outlined at the beginning of this paper. We do this by modifying a standard bar chart, noting that
\begin{itemize}
\item bar charts are familiar graphical tools,
\item the longest bar can indicate the winner,
\item each round of IRV can be depicted by a segment of the candidates' bars, and
\item ranked-choice votes can be expressed by coloring slices within those segments.
\end{itemize}

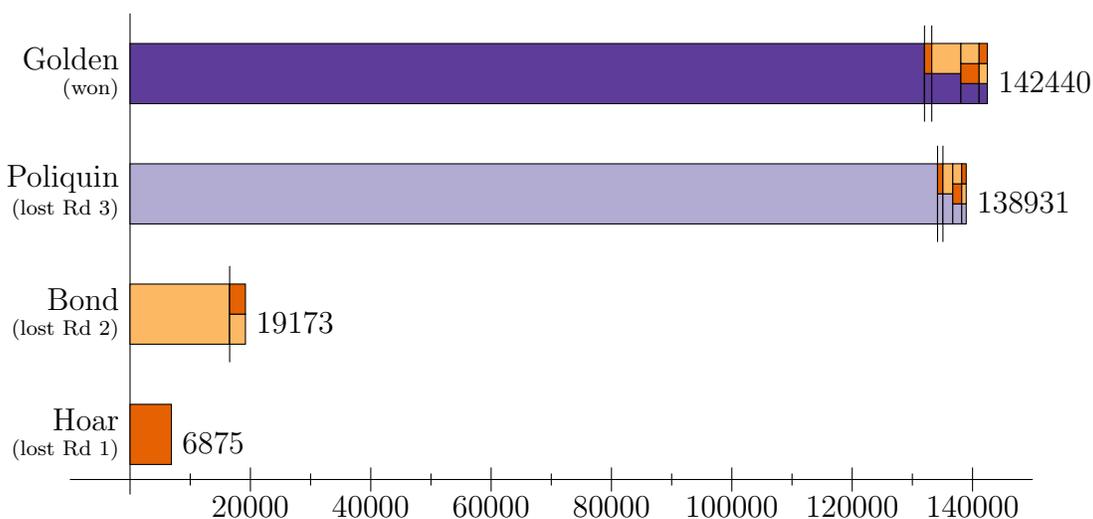
\begin{figure*}[htbp]
\begin{tikzpicture}[xscale=.08,yscale=.8]
\def \hbg {1.362}; 
\def \bhg {3.027}; 
\def \hbp {.741}; 
\def \bhp {1.485}; 
\def \bg {4.835}; 
\def \bp {1.632}; 
\def \hg {1.203}; 
\def \hp {.889}; 
\def \hb {2.621}; 
\def \hoar {6.875}; 
\def \bond {16.552}; 
\def \poliquin {134.184}; 
\def \golden {132.013}; 
%
%
\draw (0,1.5) -- (0,9.5);
\draw (0,2) coordinate (hoar);
\draw (0,4) coordinate (bond); \draw (bond)++(0,1) coordinate (bond0);
\draw (0,6) coordinate (poliquin); \draw (poliquin)++(0,1) coordinate (poliquin0);
\draw (0,8) coordinate (golden); \draw (golden)++(0,1) coordinate (golden0);

\honebar{golden}{\golden};
\honebar{poliquin}{\poliquin};
\honebar{bond}{\bond};
\honebar{hoar}{\hoar};

\hsepline{golden};
\htwobar{hoar}{golden}{\hg};
\hsepline{poliquin};
\htwobar{hoar}{poliquin}{\hp};
\hsepline{bond};
\htwobar{hoar}{bond}{\hb};

\hsepline{golden};
\htwobar{bond}{golden}{\bg};
\hthreebar{bond}{hoar}{golden}{\bhg};
\hthreebar{hoar}{bond}{golden}{\hbg};
\hsepline{poliquin};
\htwobar{bond}{poliquin}{\bp};
\hthreebar{bond}{hoar}{poliquin}{\bhp};
\hthreebar{hoar}{bond}{poliquin}{\hbp};

\draw (0,2.25) node[left] {\tiny (lost Rd 1)};
\draw (0,2.75) node[left] {Hoar};
\draw (0,4.25) node[left] {\tiny (lost Rd 2)};
\draw (0,4.75) node[left] {Bond};
\draw (0,6.25) node[left] {\tiny (lost Rd 3)};
\draw (0,6.75) node[left] {Poliquin};
\draw (0,8.25) node[left] {\tiny (won)};
\draw (0,8.75) node[left] {Golden};
\draw (-10,1.75) -- (150,1.75);
\foreach \x in {10,30,50,70,90,110,130} {\draw (\x,1.65) --++(0,.2);}
\foreach \x in {20,40,60,80,100,120,140} {\draw (\x,1.55) --++(0,.4);}
\draw (20,1.65) node[below] {$20000$};
\draw (40,1.65) node[below] {$40000$};
\draw (60,1.65) node[below] {$60000$};
\draw (80,1.65) node[below] {$80000$};
\draw (100,1.65) node[below] {$100000$};
\draw (120,1.65) node[below] {$120000$};
\draw (140,1.65) node[below] {$140000$};
\draw (golden) node[above right] {142440};
\draw (poliquin) node[above right] {138931};
\draw (bond) node[above right] {19173};
\draw (hoar) node[above right] {6875};
\end{tikzpicture}
\caption{Accumulation chart for the 2018 election in Maine's 2nd Congressional District using data from~\cite{medata}.}\label{fig:example using maine data}
\end{figure*}

We introduce accumulation charts with a demonstration --- this paper is about graphical tools, after all. This example, shown in Figure~\ref{fig:example using maine data}, describes the results of the 2018 election for Maine's 2nd Congressional District (cf.~Figure~\ref{fig:nyt maine results}).

General accumulation charts are outlined below, and we conclude this section with a second demonstration (Figure~\ref{fig:burlington - multicolored with lines}), this time using the 2009 mayoral race in Burlington, Vermont.

\setlength{\fboxsep}{10pt}
\begin{figure*}
\begin{center}
\framebox{
\begin{minipage}{5.5in}
\vspace{.1in}
\begin{center}
\underline{\textbf{Accumulation charts}}
\end{center}
\vspace{.1in}
\begin{itemize}
\item An accumulation chart is a modified bar chart, with one bar per candidate. 
\item A candidate's bar consists of multiple segments, one for each IRV round that the candidate participates in before elimination. We use long vertical lines to separate segments --- that is, to separate votes by the round in which they are accumulated.
\item The length of a candidate's Round X segment represents the number of (new) votes accumulated by the candidate during Round X. 
\item The total length of a candidate's bar represents the number of votes accumulated by that candidate in all rounds, through the candidate's final round of participation before elimination or victory. 
\item The pedigree of a vote is indicated by top-to-bottom coloring within the segment in which it gets accumulated by the candidate. In each round, we group votes by pedigree for ease of interpretation.
\end{itemize}
\vspace{.1in}
\end{minipage}}\end{center}
\end{figure*}

In order to elucidate the vertical coloring within each bar, we focus on the votes with a particular pedigree in Figure~\ref{fig:example using maine data}. Using the coloring of Figure~\ref{fig:example using maine data}, the ranked-choice vote 
\begin{center}
\begin{tabular}{|l|}
\hline
1. Hoar \ \ \ \raisebox{5pt}[15pt][0pt]{\ } \\
2. Poliquin\\
3. Bond\\
4. Golden\raisebox{5pt}[0pt][10pt]{\ }\\
\hline
\end{tabular}
\end{center}
would appear in Hoar's first-round segment as the slice
$$\begin{tikzpicture}[scale=.5]
\draw[fill=hoar-col] (0,0) rectangle (2,2);
\end{tikzpicture}$$
in Poliquin's second-round segment as the slice
$$\begin{tikzpicture}
\draw[fill=poliquin-col] (0,0) rectangle (1,.5);
\draw[fill=hoar-col] (0,.5) rectangle (1,1);
\end{tikzpicture}$$ 
not in Bond's third-round segment (because Bond was eliminated after Round 2), and in Golden's fourth-round segment (if a fourth round had been necessary in this election) as the slice
$$\begin{tikzpicture}
\draw[fill=golden-col] (0,0) rectangle (1,.25);
\draw[fill=bond-col] (0,.25) rectangle (1,.5);
\draw[fill=poliquin-col] (0,.5) rectangle (1,.75);
\draw[fill=hoar-col] (0,.75) rectangle (1,1);
\end{tikzpicture}.$$ 
The top-to-bottom coloring of a slice describes the ranked-choice vote's (initial) ranking. All slices in a candidate's accumulation bar have the same bottom color, because the candidate accumulated each of those votes when he or she was the votes' highest non-eliminated choice.

While some rounds contribute relatively skinny segments to the accumulation bars and might appear crunched in a printed version of the chart, this is not a concern if the chart is displayed electronically with zooming capability, providing both coarse and granular detail. We have implemented such an electronic version of accumulation charts at~\cite{charts}.

Note that the slices comprising a given segment may not all have the same number of colors. For example, in Figure~\ref{fig:example using maine data}, Golden's Round 3 segment
$$\begin{tikzpicture}[xscale=.08,yscale=.8]
\def \hbg {1.362}; 
\def \bhg {3.027}; 
\def \bg {4.835}; 
\draw (0,0) coordinate (golden); 
\htwobar{bond}{golden}{\bg};
\hthreebar{bond}{hoar}{golden}{\bhg};
\hthreebar{hoar}{bond}{golden}{\hbg};
\end{tikzpicture}$$
shows two three-colored slices and one two-colored slice. 
From left to right, these represent the 4835, 3027, and 1362 votes cast that began, respectively, as follows.
\begin{center}
\begin{tabular}{|l|}
\hline
1. Bond \ \ \ \raisebox{5pt}[15pt][0pt]{\ } \\
2. Golden\raisebox{5pt}[0pt][10pt]{\ }\\
\hline
\multicolumn{1}{r}{\ }
\end{tabular} \ \ 
\begin{tabular}{|l|}
\hline
1. Bond \ \ \ \raisebox{5pt}[15pt][0pt]{\ } \\
2. Hoar\\
3. Golden\raisebox{5pt}[0pt][10pt]{\ }\\
\hline
\end{tabular} \ \ 
\begin{tabular}{|l|}
\hline
1. Hoar \ \ \ \raisebox{5pt}[15pt][0pt]{\ } \\
2. Bond\\
3. Golden\raisebox{5pt}[0pt][10pt]{\ }\\
\hline
\end{tabular}
\end{center}
In each case, these votes are allotted to Golden during the third round of IRV tallying (after Hoar's elimination in Round 1 and Bond's elimination in Round 2), and so they appear in the third-round segment of Golden's bar in the accumulation chart.

Figure~\ref{fig:burlington - multicolored with lines} uses an
accumulation chart to present the results of the 2009 mayoral election
in Burlington, Vermont. A ```cleaner'' accumulation chart for that election could be obtained
by omitting Simpson and the write-in candidates, each of whom garners
relatively few votes, from the chart entirely. At the other extreme, each
write-in candidate could be displayed individually. We have chosen the
illustrated compromise of grouping the write-in candidates together, representing them by a single ``mega-round'' of elimination. This
demonstrates both how an accumulation chart looks when there are more
than four candidates, and how it looks when there are candidates who
receive few votes. The trade-offs of these choices can be explored in the interactive
accumulation charts found at~\cite{charts}.

\begin{figure*}[htbp]
\begin{tikzpicture}[xscale=3,yscale=.8]
\definecolor{smith-col}{rgb}{0.902,0.380,0.004}
\definecolor{montroll-col}{rgb}{0.992,0.722,0.388}
\definecolor{wright-col}{rgb}{0.698,0.671,0.824}
\definecolor{kiss-col}{rgb}{0.369,0.235,0.600}
\definecolor{simpson-col}{rgb}{0.868,0.628,0.868};
\definecolor{writein-col}{rgb}{0.628,0.32,0.176};
\def \kiss {2.585}; 
\def \montroll {2.063}; 
\def \simpson {.035}; 
\def \smith {1.306}; 
\def \wright {2.951}; 
\def \writein {.036}; 
\def \rk {.006};
\def \rm {.013};
\def \rh {.002};
\def \rw {.005};
\def \nk {.014};
\def \nm {.004};
\def \nh {.009};
\def \nw {.004};
\def \hk {.356};
\def \rhk {.001};
\def \nhk {.005};
\def \hnk {.014};
\def \hm {.450};
\def \nhm {.002};
\def \hnm {.021};
\def \hnrm {.001};
\def \hw {.327};
\def \nhw {.002};
\def \hnw {.005};
\def \mk {.747};
\def \rmk {.002};
\def \mrk {.002};
\def \mnk {.020};
\def \hmk {.206};
\def \mhk {.271};
\def \rmhk {.002};
\def \mrhk {.001};
\def \nhmk {.001};
\def \mnhk {.010};
\def \hnmk {.013};
\def \hmnk {.021};
\def \mhnk {.029};
\def \rmnhk {.002};
\def \rmhnk {.002};
\def \mnrhk {.001};
\def \hnrmk {.001};
\def \hnmrk {.001};
\def \mw {.424};
\def \nmw {.002};
\def \mnw{.008};
\def \hmw {.110};
\def \mhw {.182}; 
\def \nmhw {.001};
\def \mnhw {.010};
\def \hnmw {.002};
\def \hmnw {.015};
\def \mhnw {.013};
%
%
\draw (0,1.5) -- (0,13.5);
\draw (0,12) coordinate (kiss);
\draw (0,8) coordinate (montroll); \draw (montroll)++(0,1) coordinate (montroll0);
\draw (0,4) coordinate (simpson); \draw (simpson)++(0,1) coordinate (simpson0);
\draw (0,6) coordinate (smith); \draw (smith)++(0,1) coordinate (smith0);
\draw (0,10) coordinate (wright); \draw (wright)++(0,1) coordinate (wright0);
\draw (0,2) coordinate (writein); \draw (writein)++(0,1) coordinate (writein0);
\honebar{writein}{\writein};
\honebar{simpson}{\simpson};
\honebar{smith}{\smith};
\honebar{wright}{\wright};
\honebar{montroll}{\montroll};
\honebar{kiss}{\kiss};
\hsepline{smith}; 
\htwobar{writein}{smith}{\rh};
\hsepline{wright}; 
\htwobar{writein}{wright}{\rw};
\hsepline{montroll}; 
\htwobar{writein}{montroll}{\rm};
\hsepline{kiss}; 
\htwobar{writein}{kiss}{\rk};
\hsepline{smith}; 
\htwobar{simpson}{smith}{\nh};
\hsepline{wright}; 
\htwobar{simpson}{wright}{\nw};
\hsepline{montroll}; 
\htwobar{simpson}{montroll}{\nm};
\hsepline{kiss}; 
\htwobar{simpson}{kiss}{\nk};
\hsepline{wright}; 
\hthreebar{smith}{simpson}{wright}{\hnw};
\htwobar{smith}{wright}{\hw};
\hthreebar{simpson}{smith}{wright}{\nhw};

\hsepline{montroll}; 
\hfourbar{smith}{simpson}{writein}{montroll}{\hnrm};
\hthreebar{smith}{simpson}{montroll}{\hnm};
\htwobar{smith}{montroll}{\hm};
\hthreebar{simpson}{smith}{montroll}{\nhm};

\hsepline{kiss}; 
\hthreebar{smith}{simpson}{kiss}{\hnk};
\htwobar{smith}{kiss}{\hk};
\hthreebar{simpson}{smith}{kiss}{\nhk};
\hthreebar{writein}{smith}{kiss}{\rhk};
\hsepline{wright}; 
\htwobar{montroll}{wright}{\mw};
\hfourbar{montroll}{smith}{simpson}{wright}{\mhnw};
\hthreebar{montroll}{smith}{wright}{\mhw};
\hfourbar{montroll}{simpson}{smith}{wright}{\mnhw};
\hthreebar{montroll}{simpson}{wright}{\mnw};
\hfourbar{smith}{montroll}{simpson}{wright}{\hmnw};
\hthreebar{smith}{montroll}{wright}{\hmw};
\hfourbar{smith}{simpson}{montroll}{wright}{\hnmw};
\hfourbar{simpson}{montroll}{smith}{wright}{\nmhw};
\hthreebar{simpson}{montroll}{wright}{\nmw};

\hsepline{kiss}; 
\hfourbar{montroll}{smith}{simpson}{kiss}{\mhnk};
\hthreebar{montroll}{smith}{kiss}{\mhk};
\hfourbar{montroll}{simpson}{smith}{kiss}{\mnhk};
\hfivebar{montroll}{simpson}{writein}{smith}{kiss}{\mnrhk};
\hthreebar{montroll}{simpson}{kiss}{\mnk};
\hfourbar{montroll}{writein}{smith}{kiss}{\mrhk};
\hthreebar{montroll}{writein}{kiss}{\mrk};
\htwobar{montroll}{kiss}{\mk};
\hfourbar{smith}{montroll}{simpson}{kiss}{\hmnk};
\hthreebar{smith}{montroll}{kiss}{\hmk};
\hfivebar{smith}{simpson}{montroll}{writein}{kiss}{\hnmrk};
\hfourbar{smith}{simpson}{montroll}{kiss}{\hnmk};
\hfivebar{smith}{simpson}{writein}{montroll}{kiss}{\hnrmk};
\hfourbar{simpson}{smith}{montroll}{kiss}{\nhmk};
\hfivebar{writein}{montroll}{smith}{simpson}{kiss}{\rmhnk};
\hfourbar{writein}{montroll}{smith}{kiss}{\rmhk};
\hfivebar{writein}{montroll}{simpson}{smith}{kiss}{\rmnhk};
\hthreebar{writein}{montroll}{kiss}{\rmk};
\draw (0,2.25) node[left] {\tiny (lost Rd 1)};
\draw (0,2.75) node[left] {Write-ins};
\draw (0,4.25) node[left] {\tiny (lost Rd 2)};
\draw (0,4.75) node[left] {Simpson};
\draw (0,6.25) node[left] {\tiny (lost Rd 3)};
\draw (0,6.75) node[left] {Smith};
\draw (0,8.25) node[left] {\tiny (lost Rd 4)};
\draw (0,8.75) node[left] {Montroll};
\draw (0,10.25) node[left] {\tiny (lost Rd 5)};
\draw (0,10.75) node[left] {Wright};
\draw (0,12.25) node[left] {\tiny (won)};
\draw (0,12.75) node[left] {Kiss};
\draw (-.25,1.75) -- (4.32,1.75);
\foreach \x in {500,1000,1500,2000,2500,3000,3500,4000} {\draw (\x/1000,1.55) --++(0,.4); \draw (\x/1000,1.65) node[below] {$\x$};}
\draw (kiss) node[above right] {4313};
\draw (wright) node[above right] {4061};
\draw (montroll) node[above right] {2554};
\draw (smith) node[above right] {1317};
\draw (simpson) node[above right] {35};
\draw (writein) node[above right] {36 total};
\end{tikzpicture}
\caption{Accumulation chart for the 2009 mayoral election in
  Burlington, Vermont using ballot-level data
  from~\cite{bfpdata}.}\label{fig:burlington - multicolored with
  lines}
\end{figure*}
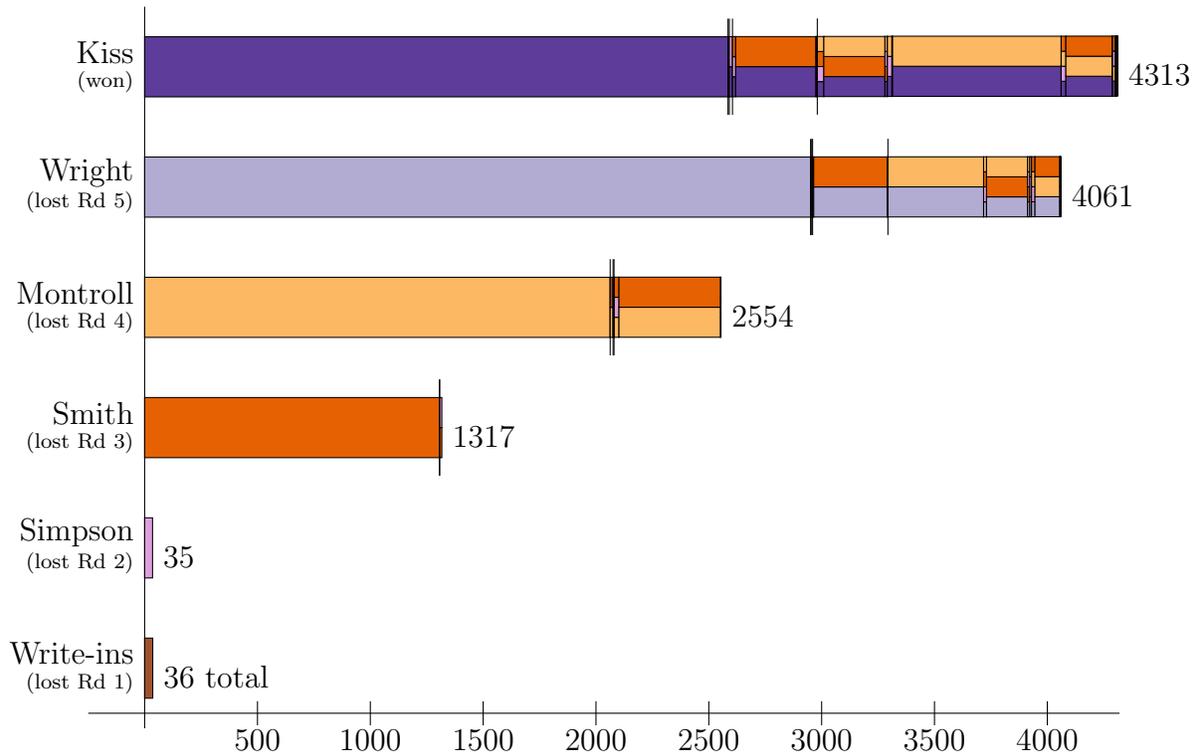

\section*{Conclusion}
\label{sec:conclusion}

Our focus in this article is on general-purpose IRV election graphics --- suitable for, say, a
newspaper. While currently there is no standard for reporting IRV results, prominent outlets have tended to present a
first-round tally followed by the final tally, such as in
Figure~\ref{fig:nyt maine results}, which neglects pertinent information.

One option for such a general purpose graphic, which has not often, to our knowledge, been used by the media for election results, is the Sankey diagram (see its use by the city of Minneapolis for its 2017 mayoral election results~\cite{mnsankey}). A Sankey diagram's primary strength is its clear depiction of how ballots are reallocated after each round. Sankey diagrams, however, fail to meet the fourth objective listed above. Moreover, we find the bands' thicknesses hard to compare in this vertical arrangement, complicating the first two objectives, as well. As such we find them less suited for the general-purpose usage we are seeking. Nonetheless, they undoubtedly provide an illuminating picture of an IRV election.

While we believe that accumulation charts successfully address the deficiencies of commonly used graphics, one must still be thoughtful about their use and analysis. For example, candidates (and their bars) get eliminated in different rounds, so they are not directly comparable. However, given the correct interpretation --- that bar length corresponds to the total number of votes accumulated before elimination --- this is not a concern. There are aspects of the election results that might be better displayed through other means (such as a Sankey diagram's depiction of vote flow) but, again, our goal here is for a general-purpose graphic. Users of these charts might choose to limit the number of rounds displayed, to order rounds from right to left in the bars, or to extend these charts by adding in information such as the number of exhausted ballots by round.

We believe that accumulation charts achieve the objectives of an election graphic. They are easy to
understand at a glance, with each candidate corresponding to a
single bar and bar lengths indicating
overall support. Importantly, accumulation charts allow a reader to follow the algorithm as it narrows
down the field, and the visible pedigrees give a sense of each candidate's coalition of support. As future elections unfold, we hope that IRV election results can be reported in such a data-rich way.


\bibliographystyle{alpha}

\end{document}